\documentclass[twocolumn,english,aps,pra,tightenlines,longbibliography]{revtex4-2}
\usepackage{lmodern}
\usepackage{amsmath}
\usepackage{amssymb}
\usepackage{graphicx}

\makeatletter
\usepackage{microtype}

\usepackage[unicode=true,pdfusetitle,
 bookmarks=true,bookmarksnumbered=false,bookmarksopen=false,
 citecolor=cyan,urlcolor=magenta,linkcolor=blue, breaklinks=false,
 pdfborder={0 0 0},backref=false,colorlinks=true]{hyperref}

\usepackage{amsthm}\usepackage{epstopdf}\usepackage{color}
\usepackage{bm}\usepackage{babel}
\usepackage{upgreek}
\usepackage{textcomp}
\usepackage{siunitx}
\newcommand\figref[2]{Fig.\,\ref{#1}\hyperref[#1]{{#2}}}

\newcommand\fref[2]{\ref{#1}\hyperref[#1]{{#2}}}

\makeatother

\usepackage{babel}
\begin{document}
\title{Tailoring sub-Doppler spectra of thermal atoms with a dielectric optical
metasurface chip }
\author{Dengke Zhang}
\email{dkzhang@buaa.edu.cn}

\author{Chen Qing}
\affiliation{School of Instrumentation and Optoelectronic Engineering, Beihang University, Beijing 100191, China}
\begin{abstract}
Compact and robust structures to precisely control and acquire atomic
spectra are increasingly important for the pursuit of widespread applications.
Sub-Doppler responses of thermal atoms are critical in constructing
high-precision devices and systems. In this study, we designed a nanograting
metasurface specifically for atomic rubidium vapor and integrated
it into a miniature vapor cell. Using the metasurface with built-in
multifunctional controls for light, we established pump-probe atomic
spectroscopy and experimentally observed sub-Doppler responses at
low incident power. Moreover, the sub-Doppler lineshape can be tailored
by varying the incident polarization state. The spectrum transformation
from absorption to transparency was observed. Using one of the sharp
responses, laser stabilization with a stability of $3\times10^{-10}$
at 2~s can be achieved. Our work reveals the effective control of
atomic spectra with optical metasurface chips, which may have great
potential for future developments in fundamental optics and novel
optical applications. 
\end{abstract}

\maketitle
\bigskip{}

\section{Introduction}

\medskip{}

Atomic-spectrum-based devices or systems such as atomic frequency
standards and clocks,\cite{Ludlow2015,Camparo2007,Diddams2004} atomic
gyroscopes,\cite{Kitching2011,Fang2012} and atomic magnetometers,\cite{Kominis2003,Budker2007}
have been widely used in various fields, including navigation and
communications,\cite{Fang2012,Bize2005} precision measurements,\cite{Hong2016,Pearman2002}
and biological imaging.\cite{Boto2018} In general, according to
the developed technology, two types of atoms are applied: the cold
atom and the thermal atomic vapor. The atomic spectra undoubtedly
play a critical role in both application technologies. To achieve
a high-performance device, precise control and acquisition of atomic
spectra are indispensable, particularly when capturing atomic responses
of hyperfine structures. Although this may not pose major challenges
for the cold atoms, intricate control and implementation processes
hinder their widespread utilization. Atom vapor is easy to obtain,
but the atomic responses are significantly broadened due to the Doppler
effect experienced by thermal atoms, which complicates the process
of capturing accurate atomic spectra and makes it difficult to achieve
the desired precision. To obtain sub-Doppler spectra, the pump-probe
configuration is typically employed using saturation absorption or
velocity selective optical pumping (VSOP).\cite{Maguire2006,Moon2008,Zigdon2009}
In addition, sub-natural responses such as electromagnetically induced
absorption (EIA) and transparency (EIT) can be observed by controlling
the state of polarization (SOP) of the pump and probe beams.\cite{Harris2006,Brazhnikov2011,Brazhnikov2018,Rehman2015,Krasteva2014,Rehman2016,Budker2002}
The SOP of light is typically adjusted using bulky optical components
of cascaded polarizers and waveplates. Thus, the optical setups for
SOP manipulation tend to be large and unstable.

In recent years, optical metasurfaces with nanostructures at subwavelength
scales have been rapidly developed and enabled control over the amplitude,\cite{Overvig2019}
phase,\cite{Kamali2018} propagation direction, \cite{Yu2011} and
polarization of transmitted and reflected light.\cite{Rubin2021}
This development has brought a range of applications such as superlenses,
holograms, sensing, and communications.\cite{Achouri2018,Chen2021}
Furthermore, optical metasurfaces have distinctive characteristics,
such as the capability to simultaneously manipulate the amplitude,
phase, and polarization, which are not attainable through conventional
optical elements.\cite{Bao2019,Overvig2020,Feng2023,Wang2023} The
emergence of nanostructured optical metasurfaces marked a significant
development in modern photonics, since these elements can supplant
or surpass the capabilities of conventional optics. Incorporating
carefully engineered nanophotonic structures can help integrate diverse
light beam manipulations into a compact chip. Consequently, a metasurface
chip can be developed to permit accurate manipulation of the intensities
and SOPs of light for the pump-probe configurations in atomic spectroscopy.
Recently, there have been several reports on the combination of metasurfaces
with thermal atoms for applications such as optical imaging,\cite{BarDavid2017}
measurement,\cite{Sebbag2021,Yang2023,Xu2023,Hummon2018} and laser
cooling,\cite{Zhu2020,Ropp2023} or theoretical investigation of
atom--surface interaction.\cite{Aljunid2016,Chan2018} However,
the regulation of atomic spectra using metasurfaces has not been extensively
explored. 

In this work, we designed a nanograting metasurface with an operating
wavelength of 780 nm, which corresponds to the D2 line of rubidium
(Rb) atoms. The optical metasurface chip was engineered to achieve
polarization-dependent reflection coefficients and phase shifts in
reflected light. In particular, the transverse magnetic (TM) polarized
light exhibits high reflectivity, whereas the transverse electric
(TE) polarized light maintains a low reflectivity. The pump-probe
configuration can be created by packaging the metasurface chip at
the rear of the vapor cell, where the incoming light serves as the
probe beam, and the reflected light functions as the pump beam. The
intensity and SOP of the pump beam can be adjusted by altering the
intensity or SOP of the incident beam. Experimentally, the sub-Doppler
spectra were observed, and its lineshape could be modified by varying
the SOP of the incident light at a low intensity. The laser stabilization
was demonstrated, and stability of $3\times10^{-10}$ at 2~s was
achieved using one of the sub-Doppler transparent responses.

\section{Results}

\subsection{Hybrid metasurface-atomic-vapor device}

\smallskip{}

\begin{figure}[ht]
\centering{}\centering \includegraphics{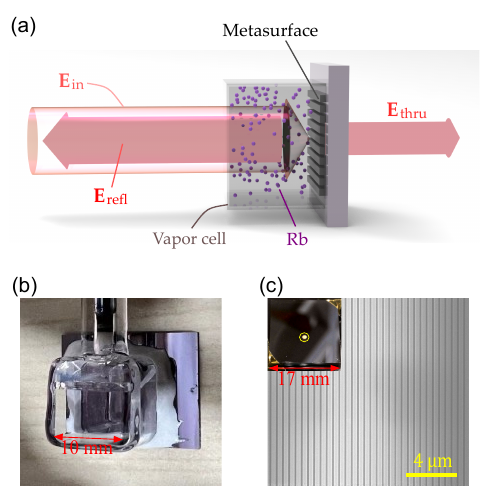} \caption{Diagram of the operation principle and photograph of the hybrid vapor
cell. a) Schematic of the vapor cell constructed by a small cubic
glass chamber, which was sealed by the designed metasurface chip.
The incident light enters the vapor cell through the glass window,
interacts with rubidium atoms, and shines on the metasurface. The
reflected light induced by the metasurface returns to the vapor and
interacts with the atoms again. A "pump-probe" configuration of
the photon--atom interaction can be built using the metasurface.
b) Photograph of a fabricated hybrid vapor cell assembled by a cubic
glass chamber and a metasurface chip. c) Scanning electron microscope
image of the grating structure of the metasurface. The inset shows
a photograph of the metasurface chip. \label{fig1}}
\end{figure}

Our hybrid metasurface-atomic-vapor device is schematically illustrated
in \figref{fig1}{a}. The device consists of an all-dielectric optical
metasurface chip, which is composed of silicon nanogratings on a borosilicate
glass substrate. A cubic quartz chamber was bonded to this metasurface
chip using a low-vapor-pressure resin sealant. Then, a rubidium (Rb)
dispenser pill was inserted, and the chamber was successively evacuated
and sealed. Finally, the Rb atoms were released by focusing a \SI{5}{\W}
semiconductor laser at \SI{915}{\nm} onto the pill for a time
duration of \SI{15}{\s}. The inner length of the cubic vapor
cell was \SI{10}{\mm}, and \figref{fig1}{b} shows a photograph
of the hybrid cell.

As shown in \figref{fig1}{a}, the light beam was injected from
the front transparent window into the vapor cell, interacted with
the thermal atoms, and scattered by the nanogratings on the back chip.
The reflected light returned to the vapor and interacted with the
atoms again. Consequently, at the normal incidence, an atom can see
both incident light and reflected light from opposite directions.
Using the birefringence and dichroism of the metasurface, the intensity
and SOP of the reflected light can be adjusted by changing the incident
light's SOP. Using this adjustable pump-probe configuration, the optical
coherence of atoms can be altered by tuning the interaction between
light and atoms, which changes their response. In principle, these
modulations will be observed in both reflectance and transmittance
spectra of the vapor cells.

\bigskip{}

\subsection{Design and characterization of metasurface}

\smallskip{}

In this hybrid vapor-cell-based pump-probe configuration, the incident
light is called the ``probe,'' and the reflected light is called
the ``pump.'' To maximize pumping, the metasurface must have high
reflectivity. Generally, metal-structured metasurfaces can have very
high reflectivity, but there is no transmitted light. The spectrum
of the probe light will help extract information of the interaction
between light and atoms. With these considerations, we adopted an
all-dielectric metasurface structure with high-contrast gratings.\cite{Karagodsky2010,ChangHasnain2012,Qiao2018}
The high-contrast gratings can offer a high reflectivity while enabling
some light to pass through the cells. In fact, by changing the SOP
of the incident light, both reflectivity and transmissivity can be
adjusted using these nanogratings. In our device, the high-contrast
gratings were constructed from silicon wires on a glass substrate.
The absorption of these thin silicon wires is weak for \SI{780}{\nm}
incident light. We performed simulations to obtain a high reflectivity
for TM polarized light using the rigorous coupled wave analysis (RCWA)
method. The gratings were designed \SI{441}{\nm} in pitch with
a duty ratio of 0.65, and the height of the silicon wires was \SI{290}{\nm}.
\figref{fig1}{c} shows a scanning electron microscope (SEM) image
of the gratings, and the inset displays the corresponding metasurface
chip, where a 1-mm-diameter circular area with the fabricated gratings
was located in the center of the chip.

\begin{figure}[ht]
\centering{}\centering \includegraphics{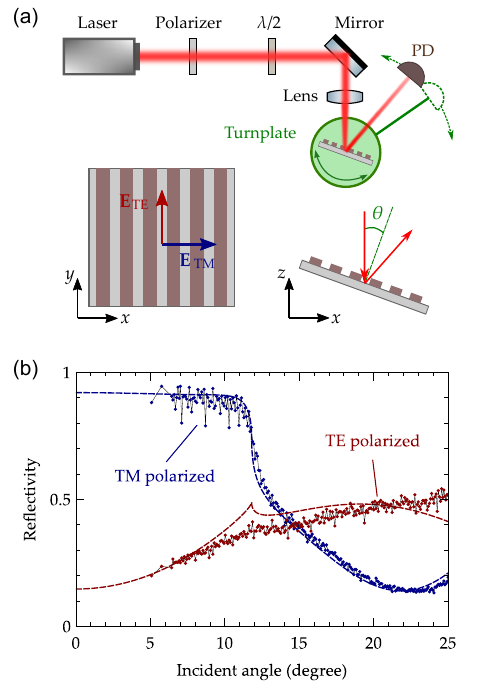} \caption{Schematic of the experimental setup and optical responses of the metasurface.
a) Experimental setup to measure the reflectivity of the metasurface
illuminated by TM and TE polarized lights at different incident angles.
$\lambda/2$: half-wave plate; PD: photodiode. The left inset sketches
the field directions of TM and TE polarized light. The right inset
describes the defined incident angle. b) Measured and simulated zeroth-order
reflectivities for both TM and TE polarized incident lights as a function
of the incident angles. The measurement was performed at $\theta>5^{\circ}$
due to the occlusion between incoming and reflected lights in the
rotating setup. The dashed curves show the simulation results, and
the dot-marker curves show the measured results. \label{fig2}}
\end{figure}

The metasurface device was characterized by measuring the reflectivity
at different incident angles. The incident light was linearly polarized
by a polarizer; then, the direction of polarization was tuned using
a half-wave plate as shown in \figref{fig2}{a}. The incident beam
finally shone on the metasurface with a focusing lens, and the reflected
beam was detected by a photodiode (PD). To vary the incident angle,
we used two electronically controlled turnplates to simultaneously
rotate the metasurface chip and PD. These all-dielectric gratings
exhibited two eigenpolarization responses, which corresponded to the
TM and TE polarized incident lights (see the left inset of \figref{fig2}{a}).
Figure \fref{fig2}{b} shows the zeroth-order reflection responses
with respect to the incident angle. The right inset of \figref{fig2}{a}
shows the definition of the incident angle $\theta$. Figure \fref{fig2}{b}
shows that the measured reflectivities for both TM and TE polarized
incident lights are consistent with the simulated results. The results
indicate that a high reflectivity of $0.93$ for TM polarized light
can be obtained at the normal incidence using the fabricated metasurface.
Meanwhile, the measured reflectance is only 0.16 for TE polarized
light. In addition, the phase difference of the reflected light between
these two polarizations is assessed by SOP measurement as $65^{\circ}$.
Therefore, it is possible to adjust the intensity and SOP of the reflected
light in a wide range by changing the SOP of the incident light.

\bigskip{}

\subsection{Theoretical analysis of the optical responses of the hybrid system}

\smallskip{}

\begin{figure*}[t]
\centering{}\centering \includegraphics{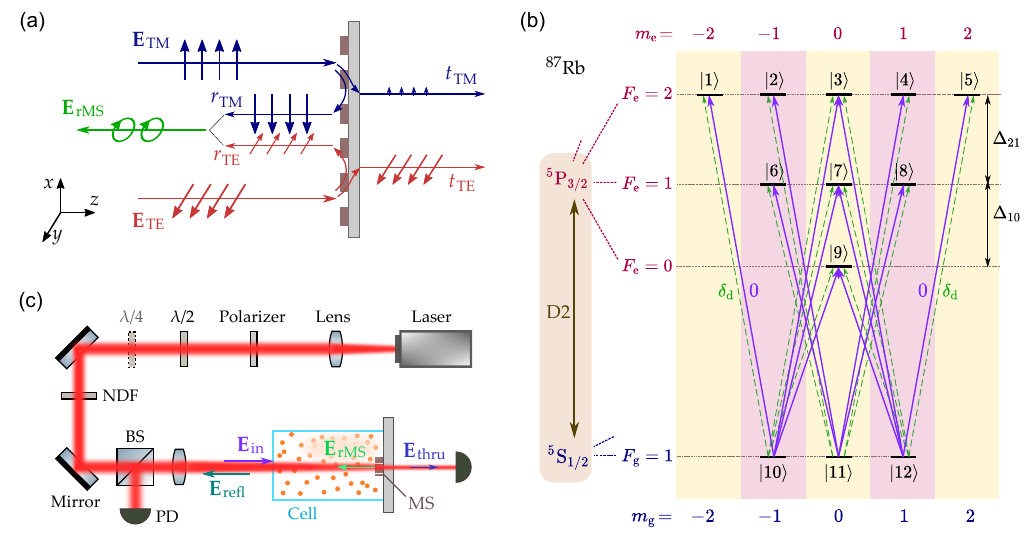} \caption{Modeling for the hybrid metasurface-atomic-vapor device and experimental
setup. a) Sketch of TM and TE polarized incident light scattered by
the metasurface. b) Energy level diagram of the D2 line of $^{87}\mathrm{Rb}$
atoms with degenerate Zeeman sublevels. The levels of $F_{\mathrm{g}}=2$
and $F_{\mathrm{e}}=3$ are not shown in this diagram. c) Schematic
description of the experimental setup. $\lambda/4$: quarter-wave
plate; NDF: neutral density filter; BS: beam splitter; MS: metasurface.
\label{fig3}}
\end{figure*}

As described for the designed metasurface, the reflection and transmission
responses for TM and TE polarized lights are different, i.e., there
is an apparent birefringence. Specifically, the intensity and SOP
of the reflected light could be dramatically adjusted by varying the
polarization of incident light. Our designed metasurface has TM and
TE polarizations as its eigenpolarizations, i.e., any polarized incident
light can be separated into two such polarized components for independent
treatment, as shown in \figref{fig3}{a}. In our assumptions, the
linearly polarized incident light was denoted by $\mathbf{E}_{\mathrm{in}}=E_{0}(\cos\phi\mathbf{\hat{e}}_{x}+\sin\phi\mathbf{\hat{e}}_{y})$,
where $E_{0}$ is the amplitude of light, $\phi$ is the polarization
angle, and $\mathbf{\hat{e}}_{x}$($\mathbf{\hat{e}}_{y}$) are the
normalized TM (TE) polarized states. Moreover, the reflected light
of the metasurface can be expressed as $\mathbf{E}_{\mathrm{rMS}}=r_{x}E_{0}\cos\phi\mathbf{\hat{e}}_{x}+r_{y}E_{0}\sin\phi\mathbf{\hat{e}}_{y}$,
where $r_{x}$($r_{y}$) are the complex reflection coefficients of
the TM (TE) polarized components. The reflected light can be transferred
to an elliptically polarized light, which is induced by the phase
difference between $r_{x}$ and $r_{y}$. Considering the interactions
of atoms with both incident and reflected lights, we selected the
propagation direction ($z$-direction) of light as the quantization
axis for atoms. For simplicity, the incident light $\mathbf{E}_{\mathrm{in}}$
and reflected light $\mathbf{E}_{\mathrm{rMS}}$ are rewritten with
circular polarization states as

\begin{equation}
\mathbf{E}_{\mathrm{in}}=\frac{E_{0}}{\sqrt{2}}\left[(\cos\phi+\mathrm{i}\sin\phi)\mathbf{\hat{e}}_{-}+(\mathrm{i}\cos\phi+\sin\phi)\mathbf{\hat{e}}_{+}\right],\label{eq:Ein}
\end{equation}

\begin{align}
\mathbf{E}_{\mathrm{\mathrm{rMS}}}= & \frac{E_{0}}{\sqrt{2}}[(r_{x}\cos\phi+\mathrm{i}r_{y}\sin\phi)\mathbf{\hat{e}}_{-}\nonumber \\
 & +(\mathrm{i}r_{x}\cos\phi+r_{y}\sin\phi)\mathbf{\hat{e}}_{+}],\label{eq:ErMS}
\end{align}
where $\mathbf{\hat{e}}_{-(+)}=[\mathbf{\hat{e}}_{x(y)}-\mathrm{i}\mathbf{\hat{e}}_{y(x)}]/\sqrt{2}$
are the spherical bases. Furthermore, $\mathbf{E}_{\mathrm{in}}$
and $\mathbf{E}_{\mathrm{\mathrm{rMS}}}$ can be redefined as $\mathbf{E}_{\mathrm{in}}=E_{0}(a_{-}\mathbf{\hat{e}}_{-}+a_{+}\mathbf{\hat{e}}_{+})/\sqrt{2}$
and $\mathbf{E}_{\mathrm{\mathrm{rMS}}}=E_{0}(b_{-}\mathbf{\hat{e}}_{-}+b_{+}\mathbf{\hat{e}}_{+})/\sqrt{2}$,
respectively. The expressions of coefficients $a_{\pm}$ and $b_{\pm}$
can be obtained using equations (\ref{eq:Ein}) and (\ref{eq:ErMS}).
In this work, we study the atomic spectra for the D2 transitions of
Rb atoms. For simplicity, we used $F_{\mathrm{g}}=1\rightarrow F_{\mathrm{e}}=0,1,2$
transitions of $^{87}\mathrm{Rb}$ atoms to explain the modeling and
computation process. Figure \fref{fig3}{b} shows the energy level
diagram of the D2 line of the $^{87}\mathrm{Rb}$ atoms. The purple
solid and green dashed arrows represent the transitions caused by
the pump and probe beams, respectively. As displayed in \figref{fig3}{b},
both $\sigma^{\pm}$ transitions can be excited by the two light beams,
since their fields are not purely circularly polarized.

In the atomic vapor, Rb atoms moving with velocity $v$ along the
light propagation direction experience the optical frequencies of
the probe and pump beams as $\omega_{\mathrm{pr}}=\omega+kv$ and
$\omega_{\mathrm{pu}}=\omega-kv$, respectively, where $\omega$ is
the optical angular frequency, and $k$ is the wave number of the
laser beam. We solve the following density matrix equation in a rotating
frame with frequency $\omega_{\mathrm{pu}}$:

\begin{equation}
{\displaystyle \rho=-\frac{\mathrm{i}}{\hbar}[H_{0}+V,\rho]+\left(\dot{\rho}\right)_{\mathrm{sp}},}\label{eq:DMequations}
\end{equation}
where $\rho$ is the density operator, $\hbar$ is the reduced Planck
constant, $H_{0}$ ($V$) is the bare atomic (interaction) Hamiltonian,
and $\left(\dot{\rho}\right)_{\mathrm{sp}}$ denotes the dissipation
process. The bare atomic Hamiltonian $H_{0}$ contains all transition-related
degenerate Zeeman sublevels, which are marked by $\left|1\right\rangle $
to $\left|12\right\rangle $ in \figref{fig3}{b}. The interaction
Hamiltonian $V$ contains the couplings of atoms with both pump and
probe beams. We can derive an effective electric susceptibility by
obtaining steady-state solutions of the density matrix elements and
averaging over the Maxwell--Boltzmann velocity distribution. For
the $\sigma^{\pm}$ components of the probe beam, the corresponding
effective electric susceptibilities are given by

\begin{align}
\chi_{\pm}= & -N_{\mathrm{Rb}}\frac{3\lambda^{3}}{4\pi^{2}}\cdot\frac{\Gamma}{\Omega_{\mathrm{0}}}\cdot\frac{1}{\sqrt{\pi}u}\sum_{F_{\mathrm{e}}=0}^{2}\sum_{m=-F_{\mathrm{g}}}^{F_{\mathrm{g}}}\frac{C_{1,m}^{F_{\mathrm{e}},m\pm1}}{a_{\pm}}\nonumber \\
 & \times\int_{-\infty}^{\infty}\mathrm{d}v\mathrm{e}^{-(v/u)^{2}}\left\langle F_{\mathrm{e}},m\pm1\left|\rho_{\mathrm{pr}}\right|F_{\mathrm{g}},m\right\rangle ,\label{eq:atomic_Chi}
\end{align}
where $\ensuremath{N_{\mathrm{Rb}}}$ is the atomic number density
of rubidium in the cell, $\Gamma$ is the decay rate of the excited
state, $\Omega_{\mathrm{0}}=\mu_{\mathrm{eg}}E_{0}/\hbar$ is the
Rabi frequency of the light beam ($\mu_{\mathrm{eg}}$: transition
dipole matrix element), $\ensuremath{u=\sqrt{2k_{\mathrm{B}}T/M}}$
is the most probable speed ($k_{\mathrm{B}}$: Boltzmann constant,
$T$: temperature of the cell, $M$: mass of an atom), $\rho_{\mathrm{pr}}$
is the density matrix elements related to interactions of the probe
beam, and $C_{F_{\mathrm{g}},m_{\mathrm{g}}}^{F_{\mathrm{e}},m_{\mathrm{e}}}$
is the normalized transition strength between the states $|F_{\mathrm{g}},m_{\mathrm{g}}\rangle$
and $|F_{\mathrm{e}},m_{\mathrm{e}}\rangle$. The detailed derivation
can be found in Section S1 in the Supporting Information. Thus, the
atomic effective refractive index is

\begin{equation}
n_{\pm}=\sqrt{1+\chi_{\pm}}.\label{eq:atomic_neff}
\end{equation}

Finally, combining the atomic response of the Rb vapor and optical
response of the metasurface, we obtain the transmitted field of the
hybrid cell:

\begin{align}
\mathbf{E}_{\mathrm{thru}}= & \frac{E_{0}t_{x}}{2}(a_{-}\mathrm{e}^{\mathrm{i}n_{-}kl_{\mathrm{c}}}-\mathrm{i}a_{+}\mathrm{e}^{\mathrm{i}n_{+}kl_{\mathrm{c}}})\mathbf{\hat{e}}_{x}\nonumber \\
 & +\frac{E_{0}t_{y}}{2}(a_{+}\mathrm{e}^{\mathrm{i}n_{+}kl_{\mathrm{c}}}-\mathrm{i}a_{-}\mathrm{e}^{\mathrm{i}n_{-}kl_{\mathrm{c}}})\mathbf{\hat{e}}_{y},\label{eq:Ethru}
\end{align}
where $t_{x}$($t_{y}$) are the complex transmission coefficients
of TM (TE) polarized lights across the metasurface, and $l_{\mathrm{c}}$
is the length of the vapor cell. Using equations (\ref{eq:Ein}--\ref{eq:Ethru}),
the atomic transmittance spectrum of the hybrid system can be obtained.
In short, when the polarization of the incident light deviates from
the eigenpolarizations of the metasurface (i.e., TM and TE polarizations),
the reflected beam will be converted into elliptically polarized light.
The following atomic transitions and populations will be simultaneously
changed. Moreover, there are multi-waves mixing and Zeeman coherences
in the multiphoton processes. As a result, the atomic polarization
and optical coherence are highly dependent on the SOP of the incident
light. Finally, the light passing through the atomic vapor will be
endowed with these features, and the effective electric susceptibility
of the vapor can be evaluated with equation (\ref{eq:atomic_Chi}).

\begin{figure*}[t]
\centering{}\centering \includegraphics{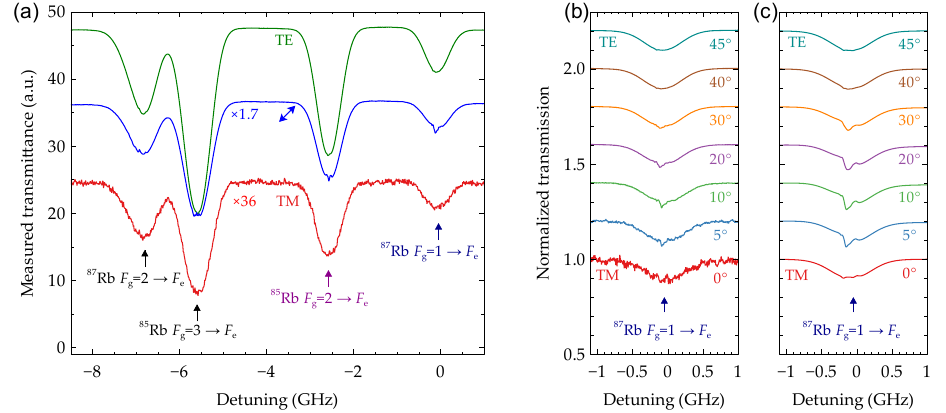} \caption{Transmittance spectra for linearly polarized incident lights. a) Experimental
power transmissions for the D2 transition of Rb atoms at TM, TE, and
$45^{\circ}$ polarized incident beams. b) Measured and c) simulated
results of normalized transmittance spectra for the transition of
$^{87}\mathrm{Rb}$ $F_{\mathrm{g}}=1$ to $F_{\mathrm{e}}=0,1,2$
for different linearly polarized incident lights. The polarization
direction of incident light is varied by changing the optical axis
angle $\vartheta$ of the half-wave plate. $\vartheta=0^{\circ}$
and $45^{\circ}$ generate TM and TE polarized lights, respectively.
In b) and c), spectrum curves for $\vartheta$ ranging from $0^{\circ}$
to $45^{\circ}$ are shown and shifted with a step of 0.2 in the $y$-axis
to improve clarity. \label{fig4}}
\end{figure*}

\bigskip{}

\subsection{Sub-Doppler atomic spectra}

\smallskip{}

Figure \fref{fig3}{c} sketches the experimental setup to measure
the transmittance and reflectance spectra of the hybrid device, where
the vapor cell was heated to $60^{\circ}\textrm{C}$. In \figref{fig3}{c},
a tunable laser operating near \SI{780}{\nm} was used to sweep
the wavelength of the incident light and linearly polarized by a polarizer.
Then, a half-wave ($\lambda/2$) plate and a quarter-wave ($\lambda/4$)
plate were used to vary the direction and ellipticity of light's polarization.
Furthermore, a neutral density filter (NDF) was utilized to attenuate
the laser beam power. With a focusing lens, the laser beams had circular
profiles with diameters of approximately \SI{0.3}{\mm} when
arriving at the metasurface, and the transmitted and reflected light
beams were detected with photodiodes. It is important to note that
weak focusing was employed to guarantee that the metasurface chip
operates with zeroth-order reflection. Narrow the beam size to 0.3
mm to make sure that most of the field was reflected in the metasurface
pattern. This adjustment can greatly minimize the side effects caused
by metasurface pattern. Furthermore, we aimed to avoid having an excessively
small beam size that would cause increased transit-time broadening.

First, we present the spectra of the linearly polarized light with
varying polarization directions, which was accomplished by adjusting
the optical axis angle $\vartheta$ of the $\lambda/2$ plate (without
using the $\lambda/4$ plate). Figure \fref{fig4}{a} shows the
measured transmission spectra for the transition of the Rb D2 line
for the cases of $0^{\circ}$ (TM), $90^{\circ}$ (TE), and $45^{\circ}$
linearly polarized incident beams with a power of \SI{5}{{\mu}W}
(corresponding light intensity of \SI{7}{\mW.cm^{-1}}). Due to the
high polarization extinction ratio of the metasurface, the transmitted
power increased when the incident beam changed from TM to TE polarization.
Higher transmitted power leads to a better signal-to-noise ratio during
the detection, which corresponds to a lower level of background noise
in the measured spectra. The typical D2 line spectra of both $^{87}\mathrm{Rb}$
and $^{85}\mathrm{Rb}$ atoms can be observed in the main shapes of
the Doppler broadening lineshape. Each line has a spectral width of
approximately \SI{570}{MHz}, which includes a Doppler broadening
width of \SI{550}{MHz}, a natural linewidth of \SI{6}{MHz},
and other broadening contributions such as a residual gas collision
broadening and a limited interaction time between Rb atoms and the
light beam. Specifically, the transmittance of the $45^{\circ}$ linearly
polarized light shows some differences compared with those of TM and
TE polarized lights, where the spectra for the hyperfine transitions
of $^{87}\mathrm{Rb}$ $F_{\mathrm{g}}=1\rightarrow F_{\mathrm{e}}$
and $^{85}\mathrm{Rb}$ $F_{\mathrm{g}}=2\rightarrow F_{\mathrm{e}}$
are clearly demonstrated. To further analyze these differences, we
focused on observing the last hyperfine transition spectra ($^{87}\mathrm{Rb}$
$F_{\mathrm{g}}=1\rightarrow F_{\mathrm{e}}$) by adjusting the polarization
direction of the incident light. Figure \fref{fig4}{b} shows the
experimental results for different $\vartheta$ values of the $\lambda/2$
plate. Figure \fref{fig4}{c} displays the corresponding simulation
results. There are obvious resemblances between simulated and measured
transmittance spectra.

As shown in \figref{fig4}{b}, for both TM and TE polarized incident
lights, the absorption spectra are similar to the spectrum of a single
beam that passed through a vapor cell. Due to the weak intensity of
the incident light, the VSOP effect is currently not functional in
the present pump-probe configuration. However, in the case of non-eigenpolarization
incidence, the optical coherence and interference between two circular
components can be observed. Since the two eigenpolarization components
have different complex reflection coefficients, the incident light
is converted into elliptically polarized light when it is not pure
TM or TE polarized. According to the characterization results of the
metasurface, the magnitudes of $r_{x}$ and $r_{y}$ are 0.93 and
0.16, respectively, where there is a phase difference of $65^{\circ}$.
In this pump-probe configuration, the left- and right-handed circular
components will not be balanced in the pump beam. The corresponding
$\sigma^{\pm}$ transitions will also be affected. Specifically, via
the transition of $F_{\mathrm{g}}\rightarrow F_{\mathrm{e}}\leq F_{\mathrm{g}}$,
the ground populations of Zeeman sublevels are tilted due to the unbalanced
pumping of the reflected lights. The induced variations of the optical
coherences can be explored by the probe beam. It is important to stress
that the transmittance spectrum is not obtained by simply multiplying
the atomic response with the incident light. Due to the linear dichroism
of the metasurface, the TM/TE polarized components have very different
transmission coefficients. The experimental results show that the
magnitude of $t_{y}$ is approximately 0.72, whereas that of $t_{x}$
is less than 0.01. Consequently, the metasurface is almost transparent
for the TE polarized light. Using equation (\ref{eq:Ethru}), we observe
that the overall transmittance is a result of the interference between
left- and right-handed circular components, which incorporate the
corresponding atomic responses (see Section S2, Supporting Information).
Through these interactions, the sub-Doppler responses linked to the
hyperfine structures can be resolved in our transmittance and reflectance
spectra. The measured transmittance spectra in \figref{fig4}{b}
show clear sub-Doppler absorption dips when $\vartheta$ is $5^{\circ}$,
$10^{\circ}$, and $20^{\circ}$. The corresponding reflectance spectra
are shown in Figure S4 in the Supporting Information, where the sub-Doppler
features can also be observed. Hence, we have shown that the polarization
direction of linear polarized light can be utilized to tune the atomic
spectra in our system.

\begin{figure}[ht]
\centering{}\centering \includegraphics{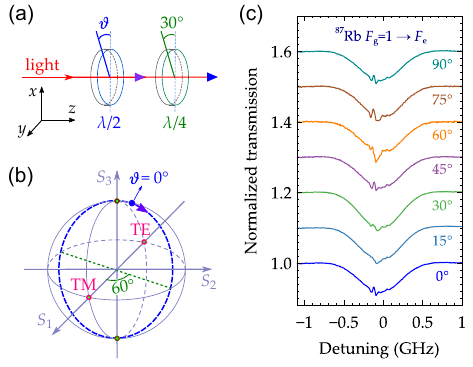} \caption{Experimental transmittance spectra for the incident lights with different
SOPs. a) The SOP of the incident light was varied using a cascaded
structure of a $\lambda/2$ plate and a $\lambda/4$ plate. In the
experiment, the optical axis of the $\lambda/4$ plate was fixed with
an angle of $30^{\circ}$, whereas the optical axis angle $\vartheta$
of the $\lambda/2$ plate was varied. b) The blue dashed circle on
the Poincar\'{e} sphere is the corresponding trace of SOP with the
operation in (a). A cycle trace will be formed for a $\pi/2$ variation
of $\vartheta$. For $\vartheta$ varying from $0^{\circ}$ to $90^{\circ}$,
the initial SOP location and evolution direction are marked on the
circle. c) Normalized transmittance spectra for the transitions of
$^{87}\mathrm{Rb}$ $F_{\mathrm{g}}=1\rightarrow F_{\mathrm{e}}$
for incident light with $\vartheta$ varying from $0^{\circ}$ to
$90^{\circ}$ in a) and b). The spectrum curves are shifted with a
step of 0.1 from $\vartheta=0^{\circ}$ to $90^{\circ}$ in the $y$-axis
to improve clarity. \label{fig5}}
\end{figure}

\bigskip{}

\subsection{Altering sub-Doppler lineshape}

\smallskip{}

To further illustrate the impact of the SOP of the incident light
beam, we adjusted the SOP of light by passing it through a cascade
of a $\lambda/2$ plate and a $\lambda/4$ plate, as shown in \figref{fig3}{c}
and \figref{fig5}{a}. Using this cascade setting, in principle,
the SOP of light can be adjusted to any state by varying the optical
axes of both plates. In our experiment, we set a fixed angle of $30^{\circ}$
for the optical axis of the $\lambda/4$ plate. The optical axis angle
$\vartheta$ of the $\lambda/2$ plate was varied to tune the SOP
of incident light. These settings help the SOP evolve from linear
polarization to circular polarization. Figure \fref{fig5}{b} shows
the corresponding SOP trace with a blue dashed circle on the Poincar\'{e}
sphere, where a cycle evolution can be achieved by a $\pi/2$ variation
of $\vartheta$. In the experiment, $\vartheta$ was varied from $0^{\circ}$
to $90^{\circ}$; Figure \fref{fig5}{b} also indicates the starting
position of the SOP and the evolution direction. Figure \fref{fig5}{c}
shows the measured results of the normalized transmittance spectra
for the transitions of $^{87}\mathrm{Rb}$ $F_{\mathrm{g}}=1\rightarrow F_{\mathrm{e}}$.
The sub-Doppler transparent peaks can be found at near circular polarizations
of the incident light, as the curves of $\vartheta=0^{\circ}$, $45^{\circ}$,
and $90^{\circ}$ in \figref{fig5}{c} show. In these cases, the
incident light is nearly circularly polarized, whereas the reflected
light is transferred to an elliptically polarized light. In other
words, the probe light interacts with Rb atoms solely through one
of the $\sigma^{\pm}$ transitions. In this pump-probe configuration,
the transmittance spectrum exhibits its sub-Doppler feature in the
transparent instead of the absorption. When the incident SOP is adjusted
from the circular polarization to a linear polarization, the sub-Doppler
transparent peaks will be converted to the sub-Doppler absorption
dip ($\vartheta=60^{\circ}$), where the maximum absorption is achieved
(see \figref{fig5}c). This evolution of the sub-Doppler lineshape
is analogous to the typical evolution for a two-path optical interference,
where the dispersion is moderated by the incident SOP incorporated
with the thermal atoms.

\begin{figure}[h]
\centering{}\centering \includegraphics{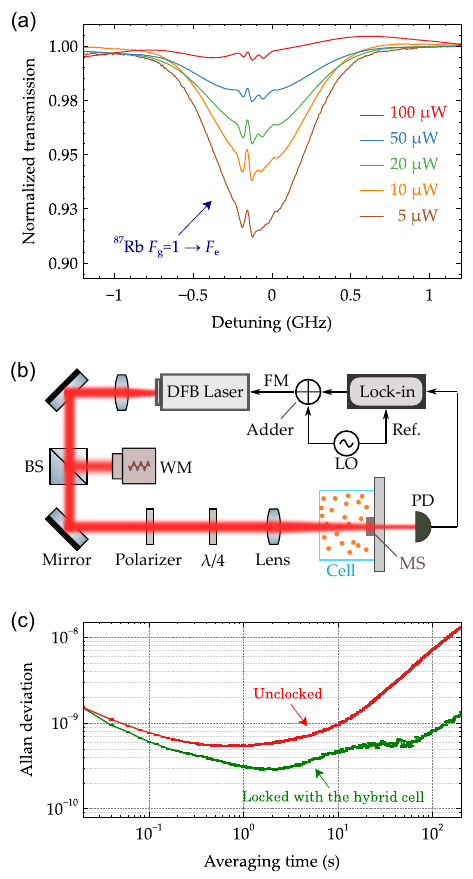} \caption{Power dependence of the transmittance spectra. a) Experimental normalized
transmittance spectra for the transitions of $^{87}\mathrm{Rb}$ $F_{\mathrm{g}}=1\rightarrow F_{\mathrm{e}}$
at different powers of the incident beam. b) Measurement setup for
the laser stabilization. The optical beams are shown in red wide lines.
Electrical connections are shown in black arrows. DFB laser: distributed
feedback laser; PA: piezoelectric actuator; WM: wavelength meter;
FM: frequency modulation; LO: local oscillator. c) Allen deviations
for the free-running DFB laser and stabilized DFB laser with the hybrid
cell. \label{fig6}}
\end{figure}

\bigskip{}

\subsection{Power dependence and laser stabilization}

\smallskip{}

To examine the effect of the beam intensity on the spectra, we performed
experiments with different powers of the incident beam in the cases
of circular polarization of the incident light. As shown in \figref{fig6}{a},
the transmission spectrum is changed by varying the power levels.
The entire attenuation induced by the absorption of atoms decreased
with the increase in power, which can be interpreted by the power
dependence saturation mechanism of two-level systems. Meanwhile, the
sub-Doppler lineshapes were altered due to the tuning of the incident
light power level, since both effective probe and pump strength were
changed. The sub-Doppler features for each hyperfine transition and
VSOP spectroscopy are displayed at the higher incident power (see
Figure S3, Supporting Information). However, there was a concurrent
decrease in interference contrast. As the five curves in \figref{fig6}{a}
show, the transparent peak has the best contrast at the power of \SI{10}{{\mu}W}.
This sharp peak can be used to stabilize lasers. To demonstrate the
stabilization performance of a laser, we performed the measurement
with the experimental setup in \figref{fig6}{b}. The distributed
feedback (DFB) laser was modulated by modulating the frequency of
the injection current. The error signal was generated using the lock-in
amplifier to demodulate the transmission signal of the hybrid vapor
cell. Then, the error signal was added to adjust the injection current
to stabilize the DFB laser. The wavelength of the DFB laser was recorded
using a high-precision wavelength meter. Figure \fref{fig6}{c}
shows the normalized Allan deviation of the measured laser wavelength.
Compared with the free-running laser, the locked laser clearly shows
better wavelength stabilization with the Allan deviation of $3\times10^{-10}$
at \SI{2}{\s}, which may be limited by the resolution of the
wavelength meter and the shifts in atomic energy levels as a background
magnetic field.

\bigskip{}

\section{Discussion and conclusion }

\medskip{}

In this work, the sub-Doppler responses for atomic hyperfine transitions
were observed using the metasurface-integrated hybrid cell in both
transmissivity and reflectivity. These sub-Doppler spectra do not
require high light intensity, which differs from the responses caused
by the VSOP effect. In fact, the VSOP effect will appear in the sub-Doppler
responses with the increase in incident power. Moreover, the sub-Doppler
lineshape in our system can be altered by adjusting the SOP of the
incident light. In the experiment, the transition from the sub-Doppler
transparent peak-to-absorption dip was observed by changing the circular
polarization to a linear polarization of the incident light. These
sub-Doppler spectra are the cooperation results of the atomic response
of the thermal atoms and the optical response of the metasurface chip
(see Section S2, Supporting Information).

To interpret the experimental spectra, we modeled this hybrid system,
and the calculations are consistent with the measured results. However,
some of the details of the sub-Doppler atomic spectra are not consistent,
and there are several possible reasons. First, for the pump light,
we did not consider the depletion of the probe light absorbed by the
atoms, especially around the resonance. This depletion will affect
the Rabi frequency of the pump beam, and the dispersion varies because
the absorption depends on the frequency. Further improvements may
be achieved by iterative calculations between optical fields and density
matrix equations. Second, the transfer of ground populations from
one hyperfine level to another is not considered. In our calculations,
we assumed that the transitions between each ground hyperfine level
to their excited states form a closed system, but this is not a true
condition in Rb atoms. The transfer of ground populations of this
type may be considered in future investigations. Third, the atom--surface
interaction was not considered. In our hybrid cell, the metasurface
structure can directly touch the atomic vapor. In this condition,
the interaction between near-field photons and atoms can occur, which
may cause further spectrum broadening due to the limited interaction
time and frequency shifts of transitions induced by the Casimir--Polder
interaction.\cite{Achouri2018,Chan2018,Stern2013,Ritter2018} In
addition, the sub-Doppler lineshape is affected by the collisions
between residual gas molecules and Rb atoms. The residual gas has
buffering and quenching functions. Collisions with the residual gas
can lead to a mixing of the Zeeman levels of the excited state, which
change the populations of the Zeeman levels. The quenching process
causes the atoms to return to their ground state without the emission
of a resonance photon. By taking into account the impact of this population
transfer between Zeeman levels and non-radiative transitions, more
accurate calculations could be achieved. In our experiment, the vacuum
level was not so good due to the limitation of the vacuum extractor.
In fact, using our metasurface chip, the spectral linewidth of sub-Doppler
responses can be reduced with a high vacuum vapor cell (see Figure
S5, Supporting Information).

In summary, we engineered a nanograting metasurface with an operating
wavelength that corresponded to the D2 line of Rb atoms. The fabricated
metasurface chip was integrated with a miniature vapor cell. By utilizing
the multifunction light control of the metasurface, we built a pump-probe
configuration in the hybrid vapor cell and observed sub-Doppler spectra
in our experiments. The lineshape of the sub-Doppler response can
be tuned by changing the SOP of the incident light at low intensity.
A spectrum transition from the absorption to the transparency was
observed. Using one of the sub-Doppler transparent responses, the
laser stabilization with $3\times10^{-10}$ instability at \SI{2}{\s}
was achieved. Our study investigated the potential applications of
optical metasurfaces in manipulating atomic spectra, which holds great
promise for future advances in fundamental optics and innovative optical
applications.

\section{Experimental Section}

\noindent \textbf{Device fabrication.} First, the all-dielectric nanograting
metasurface chip was fabricated. We used 1-mm-thickness borosilicate
glass substrate, on which the silicon film with height of \SI{290}{\nm}
was prepared by e-beam evaporation. Then the silicon nanogratings
of \SI{441}{\nm} in pitch and \SI{285}{\nm} in width were patterned
by e-beam lithography and dry etching. Second, a cubic quartz chamber
with inter length of \SI{10}{\mm} was bonded to the prepared metasurface
chip using low vapor pressure resin sealant. Then a rubidium (Rb)
dispenser pill was inserted into the chamber, which was evacuated
with vacuum degree less than \SI{0.1}{\Pa} and sealed successively.
Finally, the Rb atoms were released by focusing a \SI{915}{\nm} semiconductor
laser onto the pill for a time duration of \SI{15}{\s} at the power
of \SI{5}{\W}. The hybrid metasurface-atomic-vapor cell was fabricated
by using the above steps.

\noindent \textbf{Measurement setup.} The optical reflection of the
metasurface structure was measured with illumination of a 780-nm tunable
diode laser (Toptica DL pro) and detection of a photodiode (Thorlabs
PDA36A2). The metasurface chip and the detector were mounted on two
separated rotation stages to perform the measurements at different
incident angles. The atomic reflectance and transmittance spectra
of the hybrid cell were measured with the tunable laser and two photodiodes.
In this optical setup, waveplates, variable aperture, lenses, neutral
density filter, and beam splitters were used to adjust the polarizations,
power, and profiles of the light beams. In the experiment of laser
stabilization, a distributed feedback (DFB) laser (UniQuanta DFB801-780)
was used for the stabilization demonstration. The frequency modulated
DFB laser was controlled by adjusting the injection current with error
signal, which was generated by the demodulation of the hybrid vapor
cell transmission signal. A wavelength meter (HighFinesse WS-7) was
used to measure the wavelength of the DFB laser.

\bigskip{}

\noindent \textbf{Acknowledgements}

\noindent This work is supported by the Fundamental Research Funds
for the Central Universities under Grant KG21008401.

\medskip{}

\noindent \textbf{Conflict of Interest. }The authors declare no conflict
of interest.

\noindent \textbf{Data Availability Statement. }The data that support
the findings of this study are available from the corresponding author
upon reasonable request.

\medskip{}

\noindent \textbf{Supporting Information. }Supporting Information
is available from the Wiley Online Library or from the author.

\bibliographystyle{naturemag}

\begin{thebibliography}{10}
\expandafter\ifx\csname url\endcsname\relax
  \def\url#1{\texttt{#1}}\fi
\expandafter\ifx\csname urlprefix\endcsname\relax\def\urlprefix{URL }\fi
\providecommand{\bibinfo}[2]{#2}
\providecommand{\eprint}[2][]{\url{#2}}

\bibitem{Ludlow2015}
\bibinfo{author}{Ludlow, A.~D.}, \bibinfo{author}{Boyd, M.~M.},
  \bibinfo{author}{Ye, J.}, \bibinfo{author}{Peik, E.} \&
  \bibinfo{author}{Schmidt, P.}
\newblock \bibinfo{title}{Optical atomic clocks}.
\newblock \emph{\bibinfo{journal}{Rev. Mod. Phys.}}
  \textbf{\bibinfo{volume}{87}}, \bibinfo{pages}{637--701}
  (\bibinfo{year}{2015}).

\bibitem{Camparo2007}
\bibinfo{author}{Camparo, J.}
\newblock \bibinfo{title}{The rubidium atomic clock and basic research}.
\newblock \emph{\bibinfo{journal}{Phys. Today}} \textbf{\bibinfo{volume}{60}},
  \bibinfo{pages}{33--39} (\bibinfo{year}{2007}).

\bibitem{Diddams2004}
\bibinfo{author}{Diddams, S.~A.}, \bibinfo{author}{Bergquist, J.~C.},
  \bibinfo{author}{Jefferts, S.~R.} \& \bibinfo{author}{Oates, C.~W.}
\newblock \bibinfo{title}{Standards of time and frequency at the outset of the
  21st century}.
\newblock \emph{\bibinfo{journal}{Science}} \textbf{\bibinfo{volume}{306}},
  \bibinfo{pages}{1318--1324} (\bibinfo{year}{2004}).

\bibitem{Kitching2011}
\bibinfo{author}{Kitching, J.}, \bibinfo{author}{Knappe, S.} \&
  \bibinfo{author}{Donley, E.~A.}
\newblock \bibinfo{title}{Atomic sensors {\textendash} {{A}} review}.
\newblock \emph{\bibinfo{journal}{IEEE Sens. J.}}
  \textbf{\bibinfo{volume}{11}}, \bibinfo{pages}{1749--1758}
  (\bibinfo{year}{2011}).

\bibitem{Fang2012}
\bibinfo{author}{Fang, J.} \& \bibinfo{author}{Qin, J.}
\newblock \bibinfo{title}{Advances in atomic gyroscopes: A view from inertial
  navigation applications}.
\newblock \emph{\bibinfo{journal}{Sensors}} \textbf{\bibinfo{volume}{12}},
  \bibinfo{pages}{6331--6346} (\bibinfo{year}{2012}).

\bibitem{Kominis2003}
\bibinfo{author}{Kominis, I.~K.}, \bibinfo{author}{Kornack, T.~W.},
  \bibinfo{author}{Allred, J.~C.} \& \bibinfo{author}{Romalis, M.~V.}
\newblock \bibinfo{title}{A subfemtotesla multichannel atomic magnetometer}.
\newblock \emph{\bibinfo{journal}{Nature}} \textbf{\bibinfo{volume}{422}},
  \bibinfo{pages}{596--599} (\bibinfo{year}{2003}).

\bibitem{Budker2007}
\bibinfo{author}{Budker, D.} \& \bibinfo{author}{Romalis, M.}
\newblock \bibinfo{title}{Optical magnetometry}.
\newblock \emph{\bibinfo{journal}{Nat. Phys.}} \textbf{\bibinfo{volume}{3}},
  \bibinfo{pages}{227--234} (\bibinfo{year}{2007}).

\bibitem{Bize2005}
\bibinfo{author}{Bize, S.} \emph{et~al.}
\newblock \bibinfo{title}{Cold atom clocks and applications}.
\newblock \emph{\bibinfo{journal}{J. Phys. B: At. Mol. Opt. Phys.}}
  \textbf{\bibinfo{volume}{38}}, \bibinfo{pages}{S449--S468}
  (\bibinfo{year}{2005}).

\bibitem{Hong2016}
\bibinfo{author}{Hong, F.-L.}
\newblock \bibinfo{title}{Optical frequency standards for time and length
  applications}.
\newblock \emph{\bibinfo{journal}{Meas. Sci. Technol.}}
  \textbf{\bibinfo{volume}{28}}, \bibinfo{pages}{012002}
  (\bibinfo{year}{2016}).

\bibitem{Pearman2002}
\bibinfo{author}{Pearman, C.~P.} \emph{et~al.}
\newblock \bibinfo{title}{Polarization spectroscopy of a closed atomic
  transition: Applications to laser frequency locking}.
\newblock \emph{\bibinfo{journal}{J. Phys. B: At. Mol. Opt. Phys.}}
  \textbf{\bibinfo{volume}{35}}, \bibinfo{pages}{5141--5151}
  (\bibinfo{year}{2002}).

\bibitem{Boto2018}
\bibinfo{author}{Boto, E.} \emph{et~al.}
\newblock \bibinfo{title}{Moving magnetoencephalography towards real-world
  applications with a wearable system}.
\newblock \emph{\bibinfo{journal}{Nature}} \textbf{\bibinfo{volume}{555}},
  \bibinfo{pages}{657--661} (\bibinfo{year}{2018}).

\bibitem{Maguire2006}
\bibinfo{author}{Maguire, L.~P.}, \bibinfo{author}{van Bijnen, R. M.~W.},
  \bibinfo{author}{Mese, E.} \& \bibinfo{author}{Scholten, R.~E.}
\newblock \bibinfo{title}{Theoretical calculation of saturated absorption
  spectra for multi-level atoms}.
\newblock \emph{\bibinfo{journal}{J. Phys. B: At. Mol. Opt. Phys.}}
  \textbf{\bibinfo{volume}{39}}, \bibinfo{pages}{2709--2720}
  (\bibinfo{year}{2006}).

\bibitem{Moon2008}
\bibinfo{author}{Moon, G.} \& \bibinfo{author}{Noh, H.-R.}
\newblock \bibinfo{title}{Analytic calculation of linear susceptibility in
  velocity-dependent pump-probe spectroscopy}.
\newblock \emph{\bibinfo{journal}{Phys. Rev. A}} \textbf{\bibinfo{volume}{78}},
  \bibinfo{pages}{032506} (\bibinfo{year}{2008}).

\bibitem{Zigdon2009}
\bibinfo{author}{Zigdon, T.}, \bibinfo{author}{{Wilson-Gordon}, A.~D.} \&
  \bibinfo{author}{Friedmann, H.}
\newblock \bibinfo{title}{Absorption spectra for strong pump and probe in
  atomic beam of cesium atoms}.
\newblock \emph{\bibinfo{journal}{Phys. Rev. A}} \textbf{\bibinfo{volume}{80}},
  \bibinfo{pages}{033825} (\bibinfo{year}{2009}).

\bibitem{Harris2006}
\bibinfo{author}{Harris, M.~L.} \emph{et~al.}
\newblock \bibinfo{title}{Polarization spectroscopy in rubidium and cesium}.
\newblock \emph{\bibinfo{journal}{Phys. Rev. A}} \textbf{\bibinfo{volume}{73}},
  \bibinfo{pages}{062509} (\bibinfo{year}{2006}).

\bibitem{Brazhnikov2011}
\bibinfo{author}{Brazhnikov, D.~V.}, \bibinfo{author}{Taichenachev, A.~V.} \&
  \bibinfo{author}{Yudin, V.~I.}
\newblock \bibinfo{title}{Polarization method for controlling a sign of
  electromagnetically-induced transparency/absorption resonances}.
\newblock \emph{\bibinfo{journal}{Eur. Phys. J. D}}
  \textbf{\bibinfo{volume}{63}}, \bibinfo{pages}{315--325}
  (\bibinfo{year}{2011}).

\bibitem{Brazhnikov2018}
\bibinfo{author}{Brazhnikov, D.~V.} \emph{et~al.}
\newblock \bibinfo{title}{High-quality electromagnetically-induced absorption
  resonances in a buffer-gas-filled vapour cell}.
\newblock \emph{\bibinfo{journal}{Laser Phys. Lett.}}
  \textbf{\bibinfo{volume}{15}}, \bibinfo{pages}{025701}
  (\bibinfo{year}{2018}).

\bibitem{Rehman2015}
\bibinfo{author}{Rehman, H.-u.}, \bibinfo{author}{Adnan, M.},
  \bibinfo{author}{Noh, H.-R.} \& \bibinfo{author}{Kim, J.-T.}
\newblock \bibinfo{title}{Spectral features of electromagnetically induced
  absorption in {\textsuperscript{85}}{{Rb}} atoms}.
\newblock \emph{\bibinfo{journal}{J. Phys. B: At. Mol. Opt. Phys.}}
  \textbf{\bibinfo{volume}{48}}, \bibinfo{pages}{115502}
  (\bibinfo{year}{2015}).

\bibitem{Krasteva2014}
\bibinfo{author}{Krasteva, A.} \emph{et~al.}
\newblock \bibinfo{title}{Observation and theoretical simulation of
  electromagnetically induced transparency and enhanced velocity selective
  optical pumping in cesium vapour in a micrometric thickness optical cell}.
\newblock \emph{\bibinfo{journal}{J. Phys. B: At. Mol. Opt. Phys.}}
  \textbf{\bibinfo{volume}{47}}, \bibinfo{pages}{175004}
  (\bibinfo{year}{2014}).

\bibitem{Rehman2016}
\bibinfo{author}{Rehman, H.~U.}, \bibinfo{author}{Mohsin, M.~Q.},
  \bibinfo{author}{Noh, H.-R.} \& \bibinfo{author}{Kim, J.-T.}
\newblock \bibinfo{title}{Electromagnetically induced absorption due to
  transfer of coherence and coherence population oscillation for the
  {{F}}{\textsubscript{g}}=3{\textrightarrow{}}{{F}}{\textsubscript{e}}=4
  transition in {\textsuperscript{85}}{{Rb}} atoms}.
\newblock \emph{\bibinfo{journal}{Opt. Commun.}}
  \textbf{\bibinfo{volume}{381}}, \bibinfo{pages}{127--134}
  (\bibinfo{year}{2016}).

\bibitem{Budker2002}
\bibinfo{author}{Budker, D.} \emph{et~al.}
\newblock \bibinfo{title}{Resonant nonlinear magneto-optical effects in atoms}.
\newblock \emph{\bibinfo{journal}{Rev. Mod. Phys.}}
  \textbf{\bibinfo{volume}{74}}, \bibinfo{pages}{1153--1201}
  (\bibinfo{year}{2002}).

\bibitem{Overvig2019}
\bibinfo{author}{Overvig, A.~C.} \emph{et~al.}
\newblock \bibinfo{title}{Dielectric metasurfaces for complete and independent
  control of the optical amplitude and phase}.
\newblock \emph{\bibinfo{journal}{Light Sci. Appl.}}
  \textbf{\bibinfo{volume}{8}}, \bibinfo{pages}{92} (\bibinfo{year}{2019}).

\bibitem{Kamali2018}
\bibinfo{author}{Kamali, S.~M.}, \bibinfo{author}{Arbabi, E.},
  \bibinfo{author}{Arbabi, A.} \& \bibinfo{author}{Faraon, A.}
\newblock \bibinfo{title}{A review of dielectric optical metasurfaces for
  wavefront control}.
\newblock \emph{\bibinfo{journal}{Nanophotonics}} \textbf{\bibinfo{volume}{7}},
  \bibinfo{pages}{1041--1068.} (\bibinfo{year}{2018}).

\bibitem{Yu2011}
\bibinfo{author}{Yu, N.} \emph{et~al.}
\newblock \bibinfo{title}{Light propagation with phase discontinuities:
  Generalized laws of reflection and refraction}.
\newblock \emph{\bibinfo{journal}{Science}} \textbf{\bibinfo{volume}{334}},
  \bibinfo{pages}{333--337} (\bibinfo{year}{2011}).

\bibitem{Rubin2021}
\bibinfo{author}{Rubin, N.~A.}, \bibinfo{author}{Shi, Z.} \&
  \bibinfo{author}{Capasso, F.}
\newblock \bibinfo{title}{Polarization in diffractive optics and metasurfaces}.
\newblock \emph{\bibinfo{journal}{Adv. Opt. Photon.}}
  \textbf{\bibinfo{volume}{13}}, \bibinfo{pages}{836} (\bibinfo{year}{2021}).

\bibitem{Achouri2018}
\bibinfo{author}{Achouri, K.} \& \bibinfo{author}{Caloz, C.}
\newblock \bibinfo{title}{Design, concepts, and applications of electromagnetic
  metasurfaces}.
\newblock \emph{\bibinfo{journal}{Nanophotonics}} \textbf{\bibinfo{volume}{7}},
  \bibinfo{pages}{1095--1116} (\bibinfo{year}{2018}).

\bibitem{Chen2021}
\bibinfo{author}{Chen, Z.} \& \bibinfo{author}{Segev, M.}
\newblock \bibinfo{title}{Highlighting photonics: looking into the next
  decade}.
\newblock \emph{\bibinfo{journal}{eLight}} \textbf{\bibinfo{volume}{1}},
  \bibinfo{pages}{2} (\bibinfo{year}{2021}).

\bibitem{Bao2019}
\bibinfo{author}{Bao, Y.}, \bibinfo{author}{Ni, J.} \& \bibinfo{author}{Qiu,
  C.-W.}
\newblock \bibinfo{title}{A minimalist single-layer metasurface for arbitrary
  and full control of vector vortex beams}.
\newblock \emph{\bibinfo{journal}{Adv. Mater.}} \textbf{\bibinfo{volume}{32}},
  \bibinfo{pages}{1905659} (\bibinfo{year}{2019}).

\bibitem{Overvig2020}
\bibinfo{author}{Overvig, A.~C.}, \bibinfo{author}{Malek, S.~C.} \&
  \bibinfo{author}{Yu, N.}
\newblock \bibinfo{title}{Multifunctional {{Nonlocal Metasurfaces}}}.
\newblock \emph{\bibinfo{journal}{Phys. Rev. Lett.}}
  \textbf{\bibinfo{volume}{125}}, \bibinfo{pages}{017402}
  (\bibinfo{year}{2020}).

\bibitem{Feng2023}
\bibinfo{author}{Feng, Z.} \emph{et~al.}
\newblock \bibinfo{title}{Dual-band polarized upconversion photoluminescence
  enhanced by resonant dielectric metasurfaces}.
\newblock \emph{\bibinfo{journal}{eLight}} \textbf{\bibinfo{volume}{3}},
  \bibinfo{pages}{21} (\bibinfo{year}{2023}).

\bibitem{Wang2023}
\bibinfo{author}{Wang, S.}, \bibinfo{author}{Wen, S.}, \bibinfo{author}{Deng,
  Z.-L.}, \bibinfo{author}{Li, X.} \& \bibinfo{author}{Yang, Y.}
\newblock \bibinfo{title}{Metasurface-based solid {{Poincar\'e}} sphere
  polarizer}.
\newblock \emph{\bibinfo{journal}{Phys. Rev. Lett.}}
  \textbf{\bibinfo{volume}{130}}, \bibinfo{pages}{123801}
  (\bibinfo{year}{2023}).

\bibitem{BarDavid2017}
\bibinfo{author}{{Bar-David}, J.}, \bibinfo{author}{Stern, L.} \&
  \bibinfo{author}{Levy, U.}
\newblock \bibinfo{title}{Dynamic {{Control}} over the {{Optical Transmission}}
  of {{Nanoscale Dielectric Metasurface}} by {{Alkali Vapors}}}.
\newblock \emph{\bibinfo{journal}{Nano Lett.}} \textbf{\bibinfo{volume}{17}},
  \bibinfo{pages}{1127--1131} (\bibinfo{year}{2017}).

\bibitem{Sebbag2021}
\bibinfo{author}{Sebbag, Y.}, \bibinfo{author}{Talker, E.},
  \bibinfo{author}{Naiman, A.}, \bibinfo{author}{Barash, Y.} \&
  \bibinfo{author}{Levy, U.}
\newblock \bibinfo{title}{Demonstration of an integrated nanophotonic
  chip-scale alkali vapor magnetometer using inverse design}.
\newblock \emph{\bibinfo{journal}{Light Sci. Appl.}}
  \textbf{\bibinfo{volume}{10}}, \bibinfo{pages}{54} (\bibinfo{year}{2021}).

\bibitem{Yang2023}
\bibinfo{author}{Yang, X.}, \bibinfo{author}{Benelajla, M.},
  \bibinfo{author}{Carpenter, S.} \& \bibinfo{author}{Choy, J.~T.}
\newblock \bibinfo{title}{Analysis of atomic magnetometry using metasurface
  optics for balanced polarimetry}.
\newblock \emph{\bibinfo{journal}{Opt. Express}} \textbf{\bibinfo{volume}{31}},
  \bibinfo{pages}{13436--13446} (\bibinfo{year}{2023}).

\bibitem{Xu2023}
\bibinfo{author}{Xu, Y.} \emph{et~al.}
\newblock \bibinfo{title}{Atomic spin detection method based on spin-selective
  beam-splitting metasurface}.
\newblock \emph{\bibinfo{journal}{Adv. Optical Mater.}}
  \bibinfo{pages}{2301353} (\bibinfo{year}{2023}).

\bibitem{Hummon2018}
\bibinfo{author}{Hummon, M.~T.} \emph{et~al.}
\newblock \bibinfo{title}{Photonic chip for laser stabilization to an atomic
  vapor with 10{\textsuperscript{-11}} instability}.
\newblock \emph{\bibinfo{journal}{Optica}} \textbf{\bibinfo{volume}{5}},
  \bibinfo{pages}{443} (\bibinfo{year}{2018}).

\bibitem{Zhu2020}
\bibinfo{author}{Zhu, L.} \emph{et~al.}
\newblock \bibinfo{title}{A dielectric metasurface optical chip for the
  generation of cold atoms}.
\newblock \emph{\bibinfo{journal}{Sci. Adv.}} \textbf{\bibinfo{volume}{6}},
  \bibinfo{pages}{eabb6667} (\bibinfo{year}{2020}).

\bibitem{Ropp2023}
\bibinfo{author}{Ropp, C.} \emph{et~al.}
\newblock \bibinfo{title}{Integrating planar photonics for multi-beam
  generation and atomic clock packaging on chip}.
\newblock \emph{\bibinfo{journal}{Light Sci. Appl.}}
  \textbf{\bibinfo{volume}{12}}, \bibinfo{pages}{83} (\bibinfo{year}{2023}).

\bibitem{Aljunid2016}
\bibinfo{author}{Aljunid, S.~A.} \emph{et~al.}
\newblock \bibinfo{title}{Atomic response in the near-field of nanostructured
  plasmonic metamaterial}.
\newblock \emph{\bibinfo{journal}{Nano Lett.}} \textbf{\bibinfo{volume}{16}},
  \bibinfo{pages}{3137--3141} (\bibinfo{year}{2016}).

\bibitem{Chan2018}
\bibinfo{author}{Chan, E.~A.} \emph{et~al.}
\newblock \bibinfo{title}{Tailoring optical metamaterials to tune the
  atom-surface casimir-polder interaction}.
\newblock \emph{\bibinfo{journal}{Sci. Adv.}} \textbf{\bibinfo{volume}{4}},
  \bibinfo{pages}{eaao4223} (\bibinfo{year}{2018}).

\bibitem{Karagodsky2010}
\bibinfo{author}{Karagodsky, V.}, \bibinfo{author}{Sedgwick, F.~G.} \&
  \bibinfo{author}{Chang-Hasnain, C.~J.}
\newblock \bibinfo{title}{Theoretical analysis of subwavelength high contrast
  grating reflectors}.
\newblock \emph{\bibinfo{journal}{Opt. Express}} \textbf{\bibinfo{volume}{18}},
  \bibinfo{pages}{16973} (\bibinfo{year}{2010}).

\bibitem{ChangHasnain2012}
\bibinfo{author}{Chang-Hasnain, C.~J.} \& \bibinfo{author}{Yang, W.}
\newblock \bibinfo{title}{High-contrast gratings for integrated
  optoelectronics}.
\newblock \emph{\bibinfo{journal}{Adv. Opt. Photon.}}
  \textbf{\bibinfo{volume}{4}}, \bibinfo{pages}{379} (\bibinfo{year}{2012}).

\bibitem{Qiao2018}
\bibinfo{author}{Qiao, P.}, \bibinfo{author}{Yang, W.} \&
  \bibinfo{author}{{Chang-Hasnain}, C.~J.}
\newblock \bibinfo{title}{Recent advances in high-contrast metastructures,
  metasurfaces, and photonic crystals}.
\newblock \emph{\bibinfo{journal}{Adv. Opt. Photon.}}
  \textbf{\bibinfo{volume}{10}}, \bibinfo{pages}{180--245}
  (\bibinfo{year}{2018}).

\bibitem{Stern2013}
\bibinfo{author}{Stern, L.}, \bibinfo{author}{Desiatov, B.},
  \bibinfo{author}{Goykhman, I.} \& \bibinfo{author}{Levy, U.}
\newblock \bibinfo{title}{Nanoscale light\textendash matter interactions in
  atomic cladding waveguides}.
\newblock \emph{\bibinfo{journal}{Nat. Commun.}} \textbf{\bibinfo{volume}{4}},
  \bibinfo{pages}{1548} (\bibinfo{year}{2013}).

\bibitem{Ritter2018}
\bibinfo{author}{Ritter, R.} \emph{et~al.}
\newblock \bibinfo{title}{Coupling {{Thermal Atomic Vapor}} to {{Slot
  Waveguides}}}.
\newblock \emph{\bibinfo{journal}{Phys. Rev. X}} \textbf{\bibinfo{volume}{8}},
  \bibinfo{pages}{021032} (\bibinfo{year}{2018}).

\end{thebibliography}

\end{document}